\documentclass[twovoluments,showpacs,preprintnumbers]{revtex4}
\usepackage{amssymb}
\usepackage{amsmath}
\usepackage{txfonts}
\usepackage[mathscr]{eucal}

\begin{document}

\title{Spectroscopy of the rotating BTZ black hole via adiabatic invariance}
\author{Xian-Ming Liu$^{1,2)}$}
\affiliation{1)Department of Physics, Hubei University for Nationalities, Enshi Hubei,
445000, China; 2)Department of Physics, Institute of Theoretical Physics,
Beijing Normal University, Beijing, 100875, China}
\author{Xiao-Xiong Zeng$^{1,2)}$}
\affiliation{1)Department of Physics and Engineering Technology,
Sichuan University of Arts and Science, Dazhou Sichuan, 635000, China;
2)Department of Physics, Institute of Theoretical Physics, Beijing Normal
University, Beijing, 100875, China}
\author{Wen-Biao Liu (corresponding author)}
\email{wbliu@bnu.edu.cn}
\affiliation{Department of Physics, Institute of Theoretical Physics, Beijing Normal
University, Beijing, 100875, China}

\begin{abstract}
According to Bohr-Sommerfeld quantization rule, an equally spaced horizon
area spectrum of a static, spherically symmetric black hole was obtained
under an adiabatic invariant action. This method can be extended to the
rotating black holes. As an example, we apply this method to the rotating
BTZ black hole and obtain the quantized spectrum of the horizon area. It is
shown that the area spectrum of the rotating BTZ black hole is equally
spaced and irrelevant to the rotating parameter, which is consistent with
the Bekenstein conjecture. Specifically, the derivation do not need the
quasinormal frequencies and the small angular momentum limit.

\end{abstract}

\maketitle

\section{Introduction}

\label{Sec1}

The quantization of the black hole horizon area  has been the focus of theoretical physicists since the Bekenstein conjecture was proposed \cite
{Bekenstein1974}. Supposing that the horizon area of a black
hole behaves as a classical adiabatic invariant, Bekenstein showed that the
quantized area spectrum has the following form
\begin{equation}
(\Delta A)_{\min }=8\pi l_{p}^{2},  \label{0.1}
\end{equation}
where $l_{p}=(\frac{G\hbar }{c^{3}})^{1/2}$ is the Planck length.
Subsequently many attempts have been made to derive the area
spectrum and entropy spectrum directly utilizing the dynamical modes
of this classical
theory \cite{Bekenstein1995,Makela19961997,Dolgov1997Peleg1995,Barvinsky20012,Kastrup1996}
.

Hod \cite{Hod19982} found that if one employs the correspondence principle of Bohr, the quantized area spectrum can be determined by the real part of
quasinormal frequencies of the black hole. Hod suggested that the area spectrum is
\begin{equation}
(\Delta A)_{\min }=4\ln 3l_{p}^{2}.  \label{0.2}
\end{equation}%
Based on the Bekenstein proposal for the adiabaticity of the black hole
horizon area and and the proposal suggested by Hod regarding the quasinormal frequencies,
Kunstatter \cite{Kunstatter2003} derived the area spectrum of $d$%
-dimensional spherically symmetrical black holes. The specific result for
the horizon area quantum is same as obtained by Hod \cite{Hod19982} and by
Bekenstein and Mukhanov \cite{Bekenstein1995}.

The work done by Hod is an important step in this direction. However, Maggiore \cite%
{Maggiore2008} suggested that there are some difficulties in it.
Firstly the horizon area quantum originated from the
real part of quasinormal frequencies is not universal. Secondly,
 Hod only considered transitions from the ground state to a
state with large $n$. Once considering the general transitions
$n\rightarrow n^{\prime }$, the area changes only an arbitrarily
incremental amount. Taking the difficulties into account, Maggiore
proposed a new interpretation of the black hole quasinormal
frequencies in connection to the quantum of black hole horizon area.
Maggiore stated that a perturbed Schwarzschild black hole has to be
regarded as a damped harmonic oscillator in which the frequency should
contain both real and imaginary parts. Additionally, Maggiore showed
that the most interesting case is the highly excited quasinormal
modes in which the imaginary part is dominant compared to the real
part. The result of Maggiore in the area spectrum is also identical to
the  Eq. (\ref{0.1}) as shown by Bekenstein. Currently,
the proposal of Maggior has been widely used to study the area spectra for various black holes
\cite{ShaoWen Wei0910,Fernando2009,Deyou
Chen2010,Kwon2010,Ortega2011,Mulligan2011}.

Recently, Majhi and Vagenas \cite{Majhi and Vagenas 2011} quantized
the horizon area of a static, spherically symmetric black hole using
an adiabatic invariant $\oint pdq$ and the rule of Bohr-Sommerfeld
quantization
$\oint pdq=2\pi n$, where $p$ is the conjugate momentum of the coordinate $q$%
. The resulting black hole entropy spectrum is equally spaced and
the corresponding horizon area quantum is identical as derived by Maggiore \cite%
{Maggiore2008} and by Bekenstein \cite{Bekenstein1974}. It is interesting
that the quasinormal frequencies are not necessary to obtain the fancy
spectrum in this approach. Actually,the improved Bohr-Sommerfeld quantization rule $\oint pdq=2\pi(n+1)$
inspired by Laudau and Lifshitz has been discussed very early by Liu et al. \cite{Liu2004}. It was
applied to a Schwarzschild black hole and the area spectrum they got is $(\triangle A)=\frac{8\pi l^2_p}{3}$.
The improved quantization rule was used to explain dark matter as a remnant of black hole evaporation, where the ground state may contain an on-vanishing energy \cite{Liu2004}. However,the area spectrum is not consistent with the following references \cite{Bekenstein1995,Makela19961997,Dolgov1997Peleg1995,Barvinsky20012,Kastrup1996,Hod19982,Kunstatter2003} and
the Bekenstein conjecture \cite{Bekenstein1974}.

Let us quantize the horizon area of a rotating BTZ black hole.
It should be noted that when an adiabatic invariant is an
action variable of a classical system, the Bohr-Sommerfeld quantization
is applied as follows:
\begin{equation*}
I_{v}=\oint pdq=2\pi (n+\frac{1}{2}).
\end{equation*}%
Thus, we can quantize the horizon area of a rotating BTZ black hole
with an action variable rather than an adiabatic invariant.  It is well known that in analytical
 mechanics, the action $I$, action variable
$I_{v}$ , and the Hamiltonian $H$ of any single periodic system
should satisfy the relation
\begin{equation*}
I=I_{v}-\int {H}d\tau .
\end{equation*}%
As the action and Hamiltonian are given, the action variable can be
quantized. We find that the action variable is only the
black hole entropy, and the entropy and horizon area of a rotating
BTZ black hole can be quantized.

The remainder of this paper is arranged as follows. In Sec. \ref{Sec2}, the
rotating BTZ black hole is introduced. In the Euclidean-Kruskal form of the
rotating BTZ black hole, there is a cyclic or single periodic system with
period $\frac{2\pi }{\kappa _{+}}$, where $\kappa _{+}$ is the surface
gravity of the event horizon. In Sec. \ref{Sec3}, the area spectroscopy in
the rotating BTZ black hole is obtained.

\section{Review of a rotating BTZ black hole}

\label{Sec2}

The rotating BTZ black hole is a solution of the $(2+1)$-dimensional
Einstein gravity with a negative cosmological constant $1/l^{2}$. The
corresponding line element is \cite{BTZ1992}
\begin{equation}
ds^{2}=-N^{2}(r)dt^{2}+N^{-2}(r)dr^{2}+r^{2}(d\phi +N^{\phi }(r)dt)^{2},
\label{1.1}
\end{equation}%
where the squared lapse $N(r)$ and the angular shift $N^{\phi }(r)$ are
given as
\begin{equation}
N^{2}(r)=-M+\frac{r^{2}}{l^{2}}+\frac{J^{2}}{4r^{2}},\  \  \  \ N^{\phi }=-%
\frac{J}{2r^{2}},  \notag
\end{equation}%
with $-\infty <t<+\infty,\ 0<r<+\infty,$ and $0\leq \phi \leq 2\pi$ . Here $%
M $ and $J$ are the standard ADM mass and angular momentum of the black hole
respectively.

The rotating BTZ black hole has two horizons
\begin{equation}
r_{\pm }^{2}=\frac{1}{2}Ml^{2}(1\pm \triangle ),\  \  \  \  \triangle
=[1-(\frac{J}{Ml})^{2}]^{\frac{1}{2}},  \label{1.2}
\end{equation}%
where $r_{+},\ r_{-}$ are the outer (event) and inner (cauchy) horizon,
respectively. In particular, the existence of an event horizon means a bound on
the angular momentum $J$ as \ $|J|\leq Ml$.

For convenience, we list Hawking temperature $T_{H}$, horizon area $A_{H} $,
and angular velocity $\Omega _{H}$ at the event horizon as
\begin{equation}
T_{H}=\frac{r_{+}^{2}-r_{-}^{2}}{2\pi r_{+}l^{2}}=\frac{M\bigtriangleup }{%
2\pi r_{+}},\  \  \ A_{H}=2\pi r_{+},\  \  \  \Omega _{H}=\frac{J}{2r_{+}^{2}}.
\label{1.3}
\end{equation}%
These physical quantities can also be found in many references investigating
Hawking effect of the BTZ black hole \cite%
{Medved2002,Vagenas2002,Liu20062007,Wu and Jiang 2006,Zhang Jing Yi2010}.

To avoid the dragging effect in a rotating black hole, one should perform  the
reputed dragging coordinate transformation as
\begin{equation}
d\phi =-N^{\phi }dt=\frac{J}{2r^{2}}dt=\Omega dt,  \label{1.31}
\end{equation}%
where $\Omega $ is the dragged angular velocity of the rotating BTZ black
hole. Substituting Eq.(\ref{1.31}) into Eq.(\ref{1.1}), the line element can
be changed into the 2-dimensional form as
\begin{equation}
ds^{2}=-N^{2}(r)dt^{2}+N^{-2}(r)dr^{2}.  \label{1.4}
\end{equation}%
Thus the two-dimensional Euclidean metric reads
\begin{equation}
ds_{E}^{2}=N^{2}(r)d{\tau }^{2}+N^{-2}(r)dr^{2},  \label{1.41}
\end{equation}%
\ where we have used the Euclidean time $\tau =-it$.

After the definition of tortoise coordinate
\begin{equation*}
\frac{dr^{\ast }}{dr}=N^{-2}(r),
\end{equation*}%
or
\begin{equation*}
r^{\ast }=r+\frac{1}{2\kappa_{+}}\ln \frac{r-r_{+}}{r_{+}}-\frac{1}{2\kappa
_{-}}\ln \frac{r-r_{-}}{r_{-}},
\end{equation*}%
where $\kappa _{\pm }=\frac{M\bigtriangleup }{r_{\pm }}$ is the surface
gravity on the outer (inner) horizon, we can obtain the Euclidean Kruskal
section of the rotating BTZ black hole. Similar to the Schwarzchild black
hole, this section is a cyclic or single periodic system, whose metric reads%
\begin{equation}
ds_{E}^{2}=N^{2}(r)e^{-2\kappa_{+}r^{\ast }}(dT^{2}+dR^{2}),\  \  \  \ (r>r_{+})
\label{1.6}
\end{equation}
where
\begin{equation}
T=\frac{1}{\kappa_{+}}e^{\kappa _{+}r^{\ast }}\sin \kappa _{+}\tau,
\label{1.7}
\end{equation}%
\begin{equation}
R=\frac{1}{\kappa_{+}}e^{\kappa _{+}r^{\ast }}\cos \kappa _{+}\tau.
\label{1.8}
\end{equation}

It can been be readily seen that for both $T$ and $R$ in Eqs. (\ref{1.7}) and (\ref{1.8}) are the
periodic function of $\tau $ with period $\frac{2\pi }{\kappa _{+}}$. This
period iis critical for Hawking temperature research via the
temperature Green function\cite{Gibbons1978}. In next section, we will find
that it is also essential to derive the area and entropy spectrum of
the rotating BTZ black hole.

\section{Area spectrum of the rotating BTZ black hole}

\label{Sec3}

As is known in classical mechanics \cite{Landau1976}, one can define a
quantity called as the action variable for a single periodic system
\begin{equation*}
I_{v}=\oint pdq,
\end{equation*}%
where $p$ is the conjugate momentum of the coordinate $q$. It can be noted
that the action $I$, the action variable $I_{v}$, and the Hamiltonian ${H}$
of any single periodic system should satisfy the relation
\begin{equation}
I=I_{v}-\int {H}d\tau .  \label{2.2}
\end{equation}%
Here we prefer to use the Euclidean-like space-time Eq. (\ref{1.41}), so we
use the Euclidean time coordinate $\tau $. Based on Eq. (\ref{2.2}), it has
been found that there is a ground state energy for the Schwarzschild black
hole, which may be regarded as a candidate of dark matter \cite{Liu2004}.

Again, it can be viewed that in the dragged coordinate system, the coordinate $\phi$ does not
appear in the line element expressions as Eqs. (\ref{1.4}), (\ref{1.41}), and
(\ref{1.6}). It means that $\phi $ is an ignorable coordinate in the
Lagrangian function $L$. To eliminate this freedom completely, the action $I$
can be written as
\begin{equation}
I=\int (L-p_{\phi }\dot{\phi})d\tau =\int \int_{(0,0)}^{(p_{r},p_{\phi )}}(%
\dot{r}dp_{r}^{\prime }-\dot{\phi}dp_{\phi }^{\prime })\frac{dr}{\dot{r}},
\label{2.3}
\end{equation}%
where the dot indicates differentiation with respect to the Euclidean time $%
\tau$, $p_{r}$ and $p_{\phi }$ are the canonical momentums conjugate to $r$
and $\phi $,  respectively. This action $I$ has been used to investigate
Hawking radiation of the rotating BTZ black hole in Rfs.\cite{Wu and Jiang
2006,Zhang Jing Yi2010}. Therefore, substituting Eq.(\ref{2.3}) into Eq.(\ref%
{2.2}), the invariant action variable $I_{v}$ can be expressed as
\begin{equation}
I_{v}=\int {H}d\tau +\int \int_{(0,0)}^{(p_{r},p_{\phi )}}(\dot{r}%
dp_{r}^{\prime }-\dot{\phi}dp_{\phi }^{\prime })\frac{dr}{\dot{r}}.
\label{2.4}
\end{equation}

To remove the momentum in favor of energy, we can make use of the Hamiltonian

\begin{equation*}
\dot{r}=\frac{dH}{dp_{r}^{\prime}}\mid _{(r;\phi,p_{\phi})}=\frac{dM^{\prime}%
}{dp_{r}^{\prime}},
\end{equation*}%
\begin{equation*}
\dot{\phi }=\frac{dH}{dp_{\phi }^{\prime }}\mid _{(\phi ;r,p_{r})}=\Omega
\frac{dJ^{\prime}}{dp_{\phi }^{\prime}},
\end{equation*}%
so the invariant action variable $I_{v}$ can be rewritten as
\begin{equation}
I_{v}=\int \int_{0}^{H}d{H}^{\prime }d\tau +\int
\int_{(0,0)}^{(M,J)}[dM^{\prime }-\Omega dJ^{\prime }]\frac{dr}{\dot{r}}.
\label{2.5}
\end{equation}%
To obtain $\dot{r}$, one should consider radial, null geodesics for an
outgoing particle. This method has been used to study Hawking tunneling
radiation widely. Considering Eq.(\ref{1.41}), the radial null paths $%
(ds_{E}^{2}=0)$ can now be written as
\begin{equation}
\dot{r}\equiv \frac{dr}{d\tau }=\pm iN^{2}(r)\equiv R_{\pm }(r),
\label{2.51}
\end{equation}%
where '$\pm $' denotes the outgoing (incoming) radial null paths.
Henceforth, our subsequent analysis will focus on the outgoing paths, since
these equations are related to the quantum mechanically nontrivial features
under consideration.

Thinking of Eq.(\ref{2.51}), we have
\begin{eqnarray*}
\int \int_{0}^{H}d{H}^{\prime }d\tau &=&\int \int_{0}^{H}d{H}^{\prime }\frac{%
dr}{R_{+}(r)}=\int \int_{0}^{H}d{H}^{\prime }\frac{dr}{\dot{r}} \\
&=&\int \int_{(0,0)}^{(M,J)}[dM^{\prime }-\Omega dJ^{\prime }]\frac{dr}{\dot{%
r}},
\end{eqnarray*}%
where in the last step we have used the energy conservation condition $d{H}%
^{\prime }=dM^{\prime }-\Omega dJ^{\prime },$ and thus the adiabatic
invariant quantity given by Eq.(\ref{2.5}) should read
\begin{equation}
I_{v}=2\int \int_{0}^{H}d{H}^{\prime }d\tau .  \label{2.6}
\end{equation}

Later, we perform  the $\tau $-integration. According to the
discussion in Sec. \ref{Sec2}, we note that $\tau $ has periodicity $\frac{%
2\pi }{\kappa _{+}}.$ However, since we are considering only the outgoing
paths, the integration limit for $\tau $ will be $[0,\frac{\pi }{\kappa _{+}}%
]$ and hence the adiabatic invariant quantity reads
\begin{equation}
I_{v}=2\pi \int_{0}^{H}\frac{d{H}^{\prime }}{\kappa _{+}}.  \label{2.61}
\end{equation}

Considering $T_{H}=\frac{\kappa _{+}}{2\pi }$, the adiabatic invariant
quantity given in Eq. (\ref{2.61}) becomes

\begin{equation}
I_{v}=\int_{0}^{H}\frac{d{H}^{\prime }}{T_{H}}=S_{H},  \label{2.7}
\end{equation}%
where the first law of the rotating BTZ black hole thermodynamics has been
used.

Finally, implementing the Bohr-Sommerfeld quantization rule
\begin{equation*}
I_{v}=\oint pdq=2\pi n,
\end{equation*}%
we get the black hole entropy spectrum
\begin{equation}
S_{H}=2\pi (n+\frac{1}{2}),\  \  \  \ n=1,2,3,\ldots  \label{2.71}
\end{equation}%
Then the spacing in the entropy spectrum can be given by
\begin{equation}
\Delta S_{H}=2\pi (n+1+\frac{1}{2})-2\pi ({n+\frac{1}{2}})=2\pi.  \label{2.72}
\end{equation}%
It appears that the entropy of the rotating BTZ black hole are discrete and
the spacing is equidistant. Recalling that the black hole entropy is
proportional to the black hole horizon area(since in BTZ units $8\hbar G=1$ )
\begin{equation}
S_{H}=\frac{A}{4l_{p}^{2}},  \label{2.8}
\end{equation}
and if we employ the spacing of the entropy spectrum given in Eq.(\ref{2.72}),
the quantum of the black hole horizon area can be
\begin{equation}
\Delta A=8\pi l_{p}^{2}=\pi \hbar,  \label{2.9}
\end{equation}%
which is similar to  the Bekenstein Eq.(\ref{0.1}).

\section{Conclusions}

\label{Sec4}

With the help of Bohr-Sommerfeld quantization rule, we have derived the
quantized spectrum of the horizon area for a rotating BTZ black hole
utilizing a new approach in the context of adiabatic invariant quantities.
Our proposal is mainly based on that the Euclidean Kruskal section of the
rotating BTZ black hole is a cyclic or single periodic system. The first law
of black hole thermodynamics is also used in the derivation. It is shown
that the entropy spectrum of the rotating BTZ black hole is equally spaced
while the quantum of the horizon area identical to the results of Bekenstein.
Apparently, our result is also consistent with the initial proposal
of Bekenstein in that the area spectrum is independent on the black hole
parameters. Specifically, our analysis does not need the black hole
quasinormal frequencies, which had been used to obtain the horizon area

spectra of the rotating BTZ black hole \cite{Kwon2010}. Moreover,
the small angular momentum limit for the rotating black hole, which has been
sued by Daghigh et al. \cite{Mulligan2011} to obtain the horizon area spectra of the
Kerr black hole is not necessary here.

\begin{acknowledgments}
This research is supported by the National Natural Science Foundation of
China (Grant Nos.10773002, 10875012, 11175019). It is also supported by the
Fundamental Research Funds for the Central Universities under Grant
No.105116. Liu Xianming is also partly supported by the Team Research Program of Hubei University for Nationalities (NO. MY2011T006).
\end{acknowledgments}

\end{document}